\documentclass[preprint,epsfig,aps,prb]{revtex4}
\usepackage{graphicx}
\usepackage{latexsym}

\begin{document}
\title{Immersion microscopy based on photonic crystal materials}

\author{Igor I. Smolyaninov,$^1$ Jill Elliott$^2$, G. Wurtz$^2$, Anatoly V. Zayats,$^2$ Christopher C. Davis$^1$}
\address{$^1$Department of Electrical and Computer Engineering, University of Maryland, College
Park, MD 20742, USA}

\address{$^2$School of Mathematics and Physics, The Queen's
University of Belfast, Belfast BT7 1NN, UK}


\begin{abstract}
Theoretical model of the enhanced optical resolution of the surface plasmon immersion microscope is developed, which is based on the optics of surface plasmon Bloch waves in the tightly bound approximation. It is shown that a similar resolution enhancement may occur in a more general case of an immersion microscope based on photonic crystal materials with either positive or negative effective refractive index. Both signs of the effective refractive index have been observed in our experiments with surface plasmon immersion microscope, which is also shown to be capable of individual virus imaging.    
\end{abstract}

\maketitle  

I. INTRODUCTION.

Two-dimensional (2D) and three-dimensional (3D) photonic crystal materials have numerous applications in waveguiding and quantum optics \cite{1}. Very recently an application of these materials in far-field optical microscopy beyond the diffraction limit attracted much interest. One of the successful microscopy schemes utilizes surface plasmon polaritons (SPP) generated over the periodic nanohole array made on the gold film surface \cite{2,3}. In this 2D scheme a parabolic dielectric droplet deposited onto the gold surface behaves as a magnifying mirror for large-wavevector (short-wavelength) SPPs, which enables sub-diffraction-limited imaging. Another approach is based on the idea of a "perfect lens" made from an artificial negative refractive index metamaterial (or a thin silver film) \cite{4,5}. While both the SPP and the "perfect lens" schemes exhibit optical resolution of the order of 60 nm, which is defined by the optics of plasmon polaritons, the "perfect lens" schemes do not yet exhibit any optical magnification in the experiment (however, a theoretical scheme which exhibits magnification is described in ref. \cite{6}). Thus, at least at the moment an image formed by a "perfect lens" is observable only with an auxiliary scanning probe microscope. Since the conceptual designs of both the SPP and the "perfect lens" microscopy schemes do not look much more complicated than the design of a regular far-field optical microscope (see refs. \cite{2,3}, where the SPP microscope design is described in detail) they promise great potential in optical nanolithography and medical and biological imaging. Good understanding of the basic physics of these novel microscopy techniques is absolutely essential for making these microscopes into reliable and useful laboratory tools.   

Before we consider any real experimental geometry of a far-field optical microscope based on some 2D or 3D photonic crystal material, let us consider the spectral transmission properties of an infinite "photonic crystal space" itself. In the following simple 2D numerical example let us consider a generic object, which consists of two dots separated by a small gap (Fig.1(a)). The Fourier spectrum of this object is shown in Fig.1(b). If the spectrum of electromagnetic waves available to an experimentalist is limited by some maximum wave vector $k_{max}$ (a circle of radius $k_{max}$ is shown in Fig.1(b)), the free space behaves as a spatial frequency filter, so that whatever optical design is implemented to gather the electromagnetic waves propagating from the object, the best image which an experimentalist could get in the far-field region would result from the inverse Fourier transformation of the portion of the spectrum inside the circle in Fig.1(b). Thus, an image in Fig.1(c) would be obtained. The two-dot structure of the original object is lost in this image.

Let us now assume that the same object is placed inside an infinite periodic "photonic crystal space", while the same portion of the electromagnetic spectrum (limited by the $k_{max}$ value) is available to an experimentalist. Because of the photonic crystal periodicity, the points in Fourier space separated by integer multiples of the inverse lattice vectors are equivalent to each other. If we assume that the inverse lattice of the photonic crystal is rectangular (as shown in Fig.1(d)), the pass band of the "photonic crystal space" considered as a filter would be a rectangular lattice of circles shown in Fig.1(d). Assuming that the experimentalist is clever enough to collect all the spatial information conducted by the photonic crystal, he could get an image shown in Fig.1(e), which is obtained by the inverse Fourier transformation of the portion of the original spectrum inside all the circles. It is clear that the original information about the two-dot structure of the object may be recovered from Fig.1(e), which is also confirmed by the cross section through the image shown in Fig.1(f). 

This simple numerical example demonstrates that a far-field optical microscope with resolution beyond the $\lambda/2$ diffraction limit of usual optics can be built using photonic crystal materials. However, in order for such microscope to work, an object should be placed inside (or very near) the photonic crystal. Such a microscope should be called an immersion microscope based on photonic crystal material. It is important to emphasize that the ability to surpass the diffraction limit is not a unique property of negative refractive index metamaterials, which constitute a sub-class of a larger photonic crystal family.  
Any electromagnetic Bloch wave

\begin{equation}  
\label{eq1} 
\psi_{\vec{k}}=\sum_{\vec{K}}C_{\vec{k}-\vec{K}}e^{i(\vec{k}-\vec{K})\vec{r}},
\end{equation}

where $\vec{k}$ is defined within the first Brillouin zone, and $\vec{K}$ represents all the inverse lattice vectors, is capable of carrying spatial frequencies of an object, which would be evanescent in free space. It does not matter if the dispersion of some particular Bloch wave is negative or positive.  
What is important for microscopy, is that the Bloch waves should have sufficiently large $C_{\vec{k}-\vec{K}}$ coefficients at large $\vec{K}$. In our simple numerical example considered above this condition manifests itself as follows. If we write the Fourier spectrum $F_{\vec{\kappa}}$ of our generic object $f(\vec{r})$ in the usual way as

\begin{equation}  
\label{eq2} 
f_{\vec{r}}=\sum_{\vec{\kappa}}F_{\vec{\kappa}}e^{-i(\vec{\kappa}\vec{r}},
\end{equation}

where $\vec{\kappa}$ is understood as momentum in free space, the spectrum of the object in terms of the Bloch waves (eq.(1)) would be given as

\begin{equation}  
\label{eq3} 
F_{\vec{k}}=\sum_{\vec{K}}F_{-\vec{k}+\vec{K}}C_{\vec{k}-\vec{K}}
\end{equation}
 
Thus, high spatial frequencies of the object shape $F_{-\vec{k}+\vec{K}}$ are carried into the far-field by $C_{\vec{k}-\vec{K}}$ coefficients at large $\vec{K}$.

The limit $C_{\vec{k}-\vec{K}}\approx const$ (which is the most beneficial for superresolution imaging applications) corresponds to the tightly bound approximation, in which the photonic bands are very narrow, and in which the energy density of electromagnetic excitations (photons or plasmons) exhibits strong maxima around the lattice points of the photonic crystal. In the case of 2D microscopy based on SPP optics \cite{2,3} this Bloch wave picture is consistent with the originally suggested model of plasmon-assisted imaging using very-short-wavelength SPPs \cite{2}, which are excited by the periodic nanohole array: near the surface plasmon resonance the SPP dispersion is almost flat, which means that the tightly bound approximation is applicable. On the other hand, the tightly bound approximation itself assumes that in the limit of large lattice spacing the electromagnetic mode is tightly localized around each lattice point, which suggests a picture of some SPP mode, which is localized around some metallic motive of the photonic crystal lattice. Thus, the two models of superresolution in SPP-assisted imaging mentioned above are closely related to each other. While short-wavelength SPP picture is attractive because of its simplicity, the Bloch wave picture is free from difficulties associated with small propagation length of short-wavelength SPPs over unperturbed flat metal interface (see discussion in the next section).   

In order to be useful in far-field microscopy, a given photonic crystal geometry must exhibit image magnification to the extent that the image size should surpass the $\lambda /2$ diffraction limit of usual far-field optics. Such a magnified image can be transferred to a free space region and viewed by a regular microscope. This means that some curved photonic crystal boundary should be used. Since refraction of photonic crystals depends very strongly on frequency, propagation direction, and other parameters (a superprism effect is well-known in photonic crystal geometries, see for example \cite{7}), a reliable photonic crystal lens geometry would be very difficult to predict theoretically and realize in the experiment. On the other hand, a reflective optics geometry seems to be a good practical solution. The law of reflection is observed for almost all pseudo momenta $\vec{k}$ within the first Brillouin zone for Bloch wave reflection from a planar photonic crystal boundary (see Fig.2). The umklapp processes which occur in the corners of the Brillouin zone do not spoil the geometrical optics reflection picture because in a periodic lattice the $\vec{k}_r$ and $\vec{k}^*_r$ directions are physically equivalent, and they correspond to the same Bloch wave. If the reflecting boundary is slightly curved (so that the curvature radius is much larger than the period of the photonic crystal lattice) geometrical optics picture of reflection remains valid. Thus, a magnifying mirror can be designed using photonic crystal materials. In the 2D microscopy experiments with surface plasmon polaritons described in \cite{2,3} the role of the mirror is played by the boundary of the dielectric droplet, which is placed on the surface of the periodic nanohole array in a gold film. 

Once the image is magnified, it should be projected to free space (3D or 2D) so that it could be viewed without perturbations due to photonic crystal structure. At this stage the refractive properties of the photonic crystal play an important role in image formation. As we shall see in the following sections of this paper, the sign of the effective refractive index of the photonic crystal defines the character of image magnification of the overall optical system based on photonic crystal mirror. In these sections of the paper we will present experimental magnified images of various sub-wavelength test patterns, which were obtained in both positive and negative refractive index regimes. In addition, visualization of a T4 phage virus will be presented as an example of our technique's application to a simple biological sample.

II. THE ROLE OF EFFECTIVE REFRACTIVE INDEX OF PHOTONIC CRYSTAL MIRROR IN IMAGE MAGNIFICATION.

The 2D photonic crystal materials based on the optics of surface plasmon polaritons allow us to illustrate the behavior of the photonic crystal mirror in both positive and negative refractive index cases. A typical behavior of the SPP modes excited over a periodically modulated surface of a metal film in the vicinity of the surface plasmon resonance is shown schematically in Fig.3. Numerous examples of such behavior may be found in ref. \cite{8}, in which it is also emphasized that the imaginary part of the dielectric constant of metal does not affect the SPP Bloch modes in any considerable way. Thus, the difficulty with small propagation length of short-wavelength SPPs is alleviated. It appears that the sign of the SPP group velocity may be either positive or negative in the vicinity of the surface plasmon resonance frequency, and the SPP dispersion appears to be almost flat in this region. As we have shown above, the latter circumstance is necessary for high resolution imaging applications. On the other hand, the sign of the SPP group velocity defines the effective refractive index of the photonic crystal material in this frequency range \cite{9}. According to calculations by Kretschmann $et$ $al.$ \cite{8}, the sign of the SPP group velocity for a particular SPP branch is rather insensitive to the propagation angle, which means that the model of geometrical optics refraction (the Snell's law) is applicable to SPP propagation across the interface between the nanohole array region and the unperturbed metal film.

Let us consider the effect of this refraction on the image formation in the SPP-assisted microscope described in \cite{2,3} (see figs. 4 and 5). As may be seen from Fig.4, positive effective refractive index of the nanohole array causes some shift in the location of the image, which is formed by the photonic crystal mirror. On the other hand, negative refractive index would produce a much more drastic effect on the microscope behavior. A real image produced by the mirror, which would be located outside the nanohole array, becomes a virtual image due to negative refraction at the photonic crystal boundary (Fig.4). However, if a real image produced by the mirror is located inside the nanohole array, negative refraction at the interface produces a second real image over the unperturbed metal film (Fig.5(a)). The character of refraction at the nanohole array boundary is clearly identifiable in the experiment. While positive refraction produces image magnification which grows with distance along the optical axis of the system (this behavior has been observed in our earlier experiments \cite{2,10}), negative refraction produces an opposite behavior of magnification: image magnification is the highest in the immediate vicinity of the nanohole array boundary, and becomes smaller at larger distances along the optical axis. This may be seen from Fig.5(b), which has been calculated in the case of a triplet nanohole array, which has a negative effective refractive index (compare this figure with Figs.4 (b) and (f) from ref.\cite{2}, in which experimental ant theoretical images of a triplet nanohole array with positive refractive index were presented). Negative character of image magnification for some nanohole arrays has been indeed observed in our experiments, which are described below.   

III. EXPERIMENTAL OBSERVATIONS.

The design and operation of our SPP-assisted immersion microscope is described in detail in refs. \cite{2,3,10}. We have used a two-stage microscope design in which a magnified planar image produced originally by surface plasmon polaritons in the metal surface plane is observed with a conventional optical microscope due to SPP coupling to light via random surface roughness and nanotailored surface of the periodic nanohole array. Glycerin microdroplets have been used as 2D magnifying mirrors for SPPs. Experiments were conducted using various laser lines of an Ar-ion laser in the vicinity of 500 nm wavelength in free space, which are located very close to the frequency of the surface plasmon resonance for a gold-glycerin interface. Since the SPP wavelength and the nanohole array period are much smaller than the droplet sizes, the image formation in such a magnifying mirror can be analyzed by simple geometrical optics. To check the imaging properties of the microscope, various arrangements of periodic nano-hole arrays in a metal film were studied. Illuminated by laser light at appropriate wavelength nano-hole arrays excite the SPP modes on a structured surface of metal film \cite{11,12,13}.

Resolution test of our microscope has been performed using
30$\times$30 $\mu $m$^{2}$ arrays of singlet, doublet, and triplet nano-holes (Fig. 6) made in a gold film using focused ion beam milling (FIB) technique. These arrays exhibited positive refraction. All the test patterns shown in Fig.6 had 100 nm hole diameter with 40 nm distance between the hole edges in the doublet and triplet structures, and 500 nm lattice spacing. All the structures were clearly resolved in the optical images obtained using our SPP-assisted immersion microscope. Cross sectional analysis of these images performed in \cite{2,10} indicates spatial resolution of the order of 60 nm, which is similar to experimental resolution obtained in a "perfect lens" geometry \cite{5}. 

In the next series of experiments SPP-assisted imaging of aperiodic test samples has been studied. The aperiodic sample in Fig.7(a) was made to emulate various nanometer-scale shapes, which may appear in real-life biological applications.  Success of this experiment provides foundation for imaging of biological samples. Similar to earlier experiments, the boundary of the SPP-mirror was positioned over the aperiodic array of nanoscale holes used as objects in our imaging experiment. Illuminated with laser light, these nanoholes generate SPPs, which upon reflection from the droplet edge form magnified images of individual nanoholes over the unmodified area of the gold film (Fig. 7(b)). Zoom of the image area, which is adjacent to the square array of nanoholes (shown in Fig.7(c)) indicates that individual elements of the aperiodic array have been imaged with various degree of success (image quality appears to be the best on the right side of Fig.7(c) where the shapes of individual nanoholes are clearly recognizable). The images of the elements of the array are somewhat distorted due to their position with respect to the mirror (the same as in a conventional parabolic or elliptical mirror) and the degree of distortion is different for different elements depending on their position. Using the known mirror geometry, the shapes of the test pattern (Fig. 7(a)), and the position of the array with respect to the mirror, the distorted images of the array elements formed by the SPP mirror can be modeled and compared with the experiment (Fig. 7(d)). These images show good agreement with each other. The SPP-formed images are rotated and stretched/compressed compared to the initial structure, but the shapes of the individual elements of the array can be recognized. The additional broadening in the experimental image is related to the finite resolution of the microscope (the test pattern sizes of the order of 50 nm are comparable to the optical resolution of the apparatus), aberrations due to the imperfect mirror shape, and glycerin boundary quality.

A remarkable feature of the SPP-induced image in Fig.7(b) is the apparent inverse character of magnification in this image. This behavior is clear from the comparison of Fig.7(b) with Fig.5(a,b), and also from Fig.8 in which the distribution of magnification in this image is compared with a previously observed SPP-induced image of the triplet nanohole pattern. While in the SPP-induced image of the triplet array magnification grows with distance along the optical axis (which is consistent with a positive effective refractive index of the nanohole array), in the image of the aperiodic array magnification distribution is reversed. This behavior is consistent with negative sign of the effective refractive index of the nanohole array in Fig.7(a). Thus, both signs of the refractive index may be realized in a magnifying photonic crystal mirror.

SPP-assisted microscope has the potential to become an invaluable tool in medical and biological imaging, where far-field optical imaging of individual viruses and DNA molecules may become a reality. Water droplets on a metal surface can be used as elements of 2D optics in measurements where aqueous environment is essential for biological studies. The following experiments have been performed in order to illustrate these points. The photonic crystal mirror was used to image the T4 phage viruses (Fig. 9(a)) \cite{14} deposited onto the gold film patterned with a periodic array of doublet nanoholes.

Our initial test experiments were performed with polystyrene spheres of various diameters in the nanometer-scale range. They are widely used as test samples which emulate many features of small biological particles, such as viruses and small cells. In the experiment shown in Fig.9(b,c) 200 nm diameter polystyrene spheres obtained from Interfacial Dynamics Corporation were deposited onto a periodic nanohole array from a 0.59 nM solution in glycerin. After a 10 min exposure to this solution the sample was washed with water and methanol, and studied with a regular 3D far-field optical microscope under the illumination with 502 nm laser light. The microscope images taken in reflection (Fig.9(b)) and transmission (Fig.9(c)) indicate that individual polystyrene spheres, which have attached to the nanohole array surface as a result of our deposition process are clearly visible as standard luminous features. They appear to be very bright in the transmission image because they efficiently scatter surface plasmons into normal photons propagating in free space. On the other hand,   
because they efficiently scatter SPP energy into 3D light, they must appear as dark features in the 2D SPP-induced images. Dark features of appropriate size indeed appeared in the surface plasmon images taken with the polysterene spheres and the T4 phage viruses. The size of the scatterers deposited onto a nanohole array can be estimated from the SPP-induced image via comparison with the periodicity of the nanohole array in the image. However, the quality of the plasmon image of the nanoholes in the array appears to be somewhat worse in this case, compared to the images in Fig. 6 because of the increased plasmon and light scattering by the polystyrene spheres in glycerin. 

	Similar technique has been implemented in our experiments with the T4 phage viruses. Bacteriophages are the viruses that infect bacteria by making use of some or all of the host biosynthetic machinery. There are many similarities between bacteriophages and animal cell viruses. Thus, bacteriophages can be viewed as model systems for animal cell viruses. Direct visualization of viruses by using far-field microscopic techniques would constitute an important development in biosensing.  A typical T4 virus is around 200 nm long and 80-100 nm wide (see Fig.9(a) copied from \cite{14}). In our experiments individual T4 viruses were deposited onto an array of doublet nanoholes (shown in Fig.6). After the glycerin droplet has been placed over the array (Fig.9(d)) the plasmon image (though of reduced quality because of the scattering problems mentioned above) demonstrates resolution of the individual 100 nm nanoholes separated by 40 nm gaps, as evidenced by the image cross-sections  (Fig.9 (e) and (f)) of the plasmon image performed in the two orthogonal directions, as shown in Fig.7(e)).  In addition, the plasmon image contained quite a few dark features that are similar to the one shown in Figs.9(e,f). While the surface density of these features was consistent with the T4 phage concentration in the deposited solution, the size of these features and their appearance was consistent with the way a T4 phage should look under a microscope with 50 nm resolution. Thus, the size, image sign (T4 phages appear as dark features), and image resolution in these experiments is consistent with the known geometry of the T4 phage viruses and the resolution of the SPP-assisted microscope.

While 2D configuration of the immersion microscopy based on photonic crystal mirrors offers some important advantages, such as relative ease of photonic crystal structure fabrication, strong interaction between biological samples and SPP Bloch waves, etc., a 3D configuration of a microscope based on photonic crystal mirror is also possible. One of potentially interesting configurations is shown in Fig.10. In this configuration a photonic crystal mirror would consist of two parts: a substrate and a cover part, which would work together as a photonic crystal mirror. The plane separating these two parts should be close to the focal plane of the mirror and filled with a very thin layer of index matching gel. If an object is positioned on the substrate and covered with the top mirror part, a magnified image of the object would be formed in free space, which may be viewed by a regular optical microscope. However, practical realization of this 3D microscope idea in the optical frequency range would require fabrication of high quality 3D photonic crystal materials and very fine polishing techniques.

In conclusion, we have developed a theoretical model of the enhanced optical resolution of the surface plasmon immersion microscope, which is based on the optics of surface plasmon Bloch waves in the tightly bound approximation. It is shown that a similar resolution enhancement may occur in a more general case of an immersion microscope based on photonic crystal materials with either positive or negative effective refractive index. Both signs of the effective refractive index have been observed in our experiments with surface plasmon immersion microscope, which is also shown to be capable of individual virus imaging. 

This work has been supported in part by the NSF grant ECS-0304046 and EPSRC (UK).

\begin{figure}
\begin{center}
\end{center}
\caption{(Color on-line) (a) A generic object which consists of two dots separated by a small gap. (b) The Fourier image of the object shown in (a). An area of the spectrum of electromagnetic waves available to an experimentalist is shown by a circle. (c) The inverse Fourier image of the portion of the spectrum inside the circle in (b). (d) The pass band of the "photonic crystal space" assuming rectangular photonic crystal lattice. (e) The inverse Fourier image of the portion of the spectrum inside the photonic crystal pass band. (f) Cross section of the image in (e) indicates recovery of the two-dot structure of the object.}
\end{figure}

\begin{figure}
\begin{center}
\end{center}
\caption{(Color on-line) Reflection from a planar photonic crystal boundary satisfies the law of reflection.}
\end{figure}

\begin{figure}
\begin{center}
\end{center}
\caption{Schematic view of the SPP dispersion law in the first Brillouin zone.}
\end{figure}

\begin{figure}
\begin{center}
\end{center}
\caption{(Color on-line) Effect of the sign of the effective refractive index of the nanohole array on image formation: positive refractive index causes small shift in image location, while negative refractive index converts real images outside the nanohole array into virtual ones.}
\end{figure}

\begin{figure}
\begin{center}
\end{center}
\caption{(Color on-line) (a) Negative refractive index produces reverse distribution of magnification in images produced by the photonic crystal mirror. (b) Theoretical image of a triplet array of nanoholes in the case of negative effective refractive index of the array.}
\end{figure}

\begin{figure}
\begin{center}
\end{center}
\caption{(Color on-line) Resolution test of the microscope. The
30$\times$30 $\mu $m$^{2}$ arrays of singlet, doublet, and triplet nano-holes (100 nm hole diameter, 40 nm distance between the hole edges in the doublet and triplet, 500 nm period) shown in the left column are imaged using a SPP mirror
formed by a glycerine droplet. The optical images in the right column are obtained at $\lambda_0$ = 502 nm (singlets and doublets) and at $\lambda_0$ = 515 nm (triplets).}
\end{figure}

\begin{figure}
\begin{center}
\end{center}
\caption{(Color on-line) (a) SEM image of the aperiodic array in the metal film. (b) Large-scale image of the array obtained with SPP-assisted microscope. Total size of the array is 20x20 $\mu m^2$. Droplet edge position is shown by the dashed line. (c) Zoom into the image in (b) shown together with the calculated image of the array (d).}
\end{figure}

\begin{figure}
\begin{center}
\end{center}
\caption{(Color on-line) Comparison of magnification distribution in the SPP-induced image of the aperiodic array (a) and in the image of the triplet array (b). (c) and (d) show cross-sections of the SPP-induced images. The cross section directions are shown in (a) and (b), respectively. The corresponding plots of feature sizes in the images are shown in (e) and (f).}
\end{figure}

\begin{figure}
\begin{center}
\end{center}
\caption{(Color on-line) (a) Electron microscope image of the T4 phage. (d)  Droplet used in the imaging. (b,c) Regular microscope images of the nanohole array after polystyrene spheres deposition in reflection (b) and transmission (c) under illumination with 502 nm light. Individual polystyrene spheres appear to be very bright in transmission because they efficiently scatter surface plasmons into normal photons propagating in free space. On the other hand, T4 phage particles appear as dark features (indicated by an arrow) in the 2D plasmon image (e) and (f). Orthogonal cross-sections (g) and (h) indicate resolving of the doublets in the SPP-induced image (e).}
\end{figure}

\begin{figure}
\begin{center}
\end{center}
\caption{ Schematic view of a 3D magnifying photonic crystal mirror, which may be used in a 3D configuration of an immersion microscope based on photonic crystal materials.}
\end{figure}

\end{document}